# Self-Induced Quasistationary Magnetic Fields

E. O. Kamenetskii

Department of Electrical and Computer Engineering,

Ben Gurion University of the Negev, Beer Sheva, 84105, Israel

January 15, 2006

**Abstract**

The interaction of electromagnetic radiation with temporally dispersive magnetic solids of small dimensions may show very special resonant behaviors. The internal fields of such samples are characterized by magnetostatic-potential scalar wave functions. The oscillating modes have the energy orthogonality properties and unusual pseudo-electric (gauge) fields. Because of a phase factor, that makes the states single valued, a persistent magnetic current exists. This leads to appearance of an eigen-electric moment of a small disk sample. One of the intriguing features of the mode fields is dynamical symmetry breaking.

## 1. Introduction

For a localized region with a finite-space domain of charge distribution, standard equations of electrostatics, $\nabla \times \vec{E}(\vec{r}) = 0$; $\nabla \cdot \vec{D}(\vec{r}) = 4\pi\rho(\vec{r})$, may lead to appearance of so-called *self-induced electrostatic fields* [1]. Such fields take place when charge distributions are functions of the electrostatic potential. The basic equations for the electrostatic potential are $\Delta\phi(\vec{r}) = 0$ outside a domain of charge distribution and $\varepsilon\Delta\phi(\vec{r}) = -4\pi\rho(\phi(\vec{r}))$ inside a domain of charge distribution. Electrostatic fields generated by sources $\rho$ such that $\rho(\phi(\vec{r}) = 0) = 0$ are called as the self-induced electrostatic fields. In linear approximation (with respect to $\phi$, not with respect to $\vec{r}$) the charge density for self-induced fields are of the form $\rho(\vec{r}) = \Omega(\vec{r})\phi(\vec{r})$, where $\Omega(\vec{r})$ is some structure function specific to the particular distribution of charge in space. For a certain case, the basic equation of electrostatics of self-induced fields takes the form of the Schrödinger-type equation (stationary-state) with the quantized permittivities corresponding to discrete potential-egenfunction states [1].

The self-induced electric fields in small samples considered by Kapuścik [1] are pure static fields. Recently, the readers' attention was called to the paper by Fredkin and Mayergoyz [2], which addresses the nature of the electrostatic *resonance* behavior in small (compared to the free-space electromagnetic-wave wavelength) dielectric objects. These resonances take place for negative quantities of the temporally dispersive permittivity: $\varepsilon(\omega) < 0$. For resonance quasielectrostatic modes, there are resonance values of permittivity $\varepsilon$. For the case considered by Fredkin and Mayergoyz, quasielectrostatic fields are the fields with localized sources induced by the electrostatic potential. In the sense perceived by Kapuścik, these are, in fact, the *self-induced quasielectrostatic fields*.

It is well known that in a general case of small (compared to the free-space electromagnetic-wave wavelength) samples made of media with strong temporal dispersion, the role of displacement currents in Maxwell equations can be negligibly small, so oscillating fields are the quasistationary fields [3]. For the case considered by Fredkin and Mayergoyz, one neglects a magnetic displacement current and has quasistationary electric fields. A dual situation (with respect to quasielectrostatic resonances) is demonstrated for quasistationary magnetic fields in small samples with strong temporal dispersion of the permeability tensor: $\vec{\vec{\mu}} = \vec{\vec{\mu}}(\omega)$. In such small samples, variation of the electric energy is negligibly small compared to variation of the magnetic energy and so one can neglect the electric displacement current in Maxwell equations [3]. These magnetic samples can exhibit the magnetostatic resonance behavior in microwaves [4 – 7]. For resonance modes, there are resonance values of permeability $\mu$. So one may call these modes as the *self-induced quasimagnetostatic fields*.

When one neglects the displacement currents, one can introduce a notion of a scalar potential: electrostatic potential $\phi$ for quasielectrostatic fields and magnetostatic potential $\psi$ for quasimagnetostatic fields. These potentials, however, do not have the same physical meaning as in a situation of pure electrostatics and magnetostatics. Since there are resonant behaviors of small dielectric/magnetic objects (confinement phenomena plus temporal-dispersion conditions of tensors $\vec{\vec{\varepsilon}}(\omega)$ and/or $\vec{\vec{\mu}}(\omega)$), we have scalar *wave functions*: electrostatic-potential wave function $\phi(\vec{r},t)$ and magnetostatic-potential wave function $\psi(\vec{r},t)$. The main note is that since we are on the level of the continuum description of media (based on tensors $\vec{\vec{\varepsilon}}(\omega)$ and/or $\vec{\vec{\mu}}(\omega)$), the boundary conditions should be imposed on the scalar wave functions $\phi(\vec{r},t)$ and/or $\psi(\vec{r},t)$ and their derivatives, but not on the RF functions of polarization (plasmons) and/or magnetization (magnons). There are no electron-motion equations in the continuum ($\vec{\vec{\varepsilon}}$- and $\vec{\vec{\mu}}$-based) description.

It is clear that the quasistationary (time-variable) electric field should be accompanied with the RF magnetic field. Similarly, the quasistationary (time-variable) magnetic field should be accompanied with the RF electric field. Use of the notion of the scalar wave functions leads, however, to evident contradictions with the dynamical Maxwell equations (DME). This fact can be perceived, in particular, from the remarks made by McDonalds [8]. Let us consider a case of electrostatic resonances in small dielectric (non-conducting) objects. We introduce electrostatic potential $\phi(\vec{r},t)$ when we neglect the magnetic displacement current: $\frac{\partial \vec{B}}{\partial t} = 0$. From the Maxwell equation (the Ampere-Maxwell law), $\nabla \times \vec{H} = \frac{1}{c}\frac{\partial \vec{D}}{\partial t}$, one can write

$$\nabla \times \frac{\partial \vec{H}}{\partial t} = \frac{1}{c}\frac{\partial^2 \vec{D}}{\partial t^2}. \tag{1}$$

If a sample does not posses any magnetic anisotropy, we have

$$\frac{\partial^2 \vec{D}}{\partial t^2} = 0. \tag{2}$$

Similarly, for magnetostatic resonances in small magnetic objects one neglects the electric displacement current: $\frac{\partial \vec{D}}{\partial t} = 0$. From Maxwell equation (the Faraday law), $\nabla \times \vec{E} = -\frac{1}{c}\frac{\partial \vec{B}}{\partial t}$, one obtains

$$\nabla \times \frac{\partial \vec{E}}{\partial t} = -\frac{1}{c}\frac{\partial^2 \vec{B}}{\partial t^2}. \tag{3}$$

If a sample does not posses any dielectric anisotropy, we have

$$\frac{\partial^2 \vec{B}}{\partial t^2} = 0. \tag{4}$$

As it follows from Eqs. (2) and (4), the electric field in small resonant dielectric objects as well as the magnetic field in small resonant magnetic objects vary linearly with time. This leads, however, to arbitrary large fields at early and late times, and is excluded on physical grounds. An evident conclusion suggests itself at once: the electric (for electrostatic resonances) and magnetic (for magnetostatic resonances) fields are constant quantities. This contradicts, however, to the fact of temporally dispersive media and any resonant conditions. Another conclusion is more unexpected: for a case of electrostatic resonances the Ampere-Maxwell law is not valid and for a case of magnetostatic resonances the Faraday law is not valid. The purpose of this paper is to demonstrate that self-induced quasistationary fields of magnetostatic (MS) resonance modes in small samples are rather the Schrödinger-like (or even the Dirac-like) fields than the Maxwell-like fields. MS oscillations in small objects are characterized by the pseudo-electric (gauge) fields.

## 2. The power flow density for propagating quasistationary magnetic modes

MS ferromagnetism has a character essentially different from exchange ferromagnetism [9, 10]. This statement finds strong confirmation in confinement phenomena of magnetic-dipolar oscillations. The dipole interaction provides us with a long-range mechanism of interaction, where a magnetic medium is considered as a continuum. Contrary to an exchange spin wave, in magnetic-dipolar waves the local fluctuation of magnetization does not propagate due to interaction between the neighboring spins. There should be certain propagating fields – the MS fields – which cause and govern propagation of magnetization fluctuations. In other words, space-time magnetization fluctuations are corollary of the propagating MS fields, but there are no magnetization waves. The boundary conditions should be imposed on the MS field and not on the RF magnetization. This is slightly akin physics of propagation of electromagnetic waves in a transmission-line system. In this case the electromagnetic-wave propagation causes space-time fluctuations of a conductivity current in metal parts of a line, but there are no electric-charge-density waves. So the boundary conditions are imposed on the electromagnetic field components and not on the RF currents.

When field differences across the sample become comparable to the bulk demagnetizing fields the local-oscillator approximation is no longer valid, and indeed under certain circumstances, entirely new spin dynamics behavior can be observed. This dynamics behavior is the following.

Precession of magnetization about a vector of a bias magnetic field produces a small oscillating magnetization $\vec{m}$ and a resulting dynamic demagnetizing field $\vec{H}$, which reacts back on the precession, raising the resonant frequency. Vectors $\vec{H}$ and $\vec{m}$ are coupled by the differential relation:

$$\nabla \cdot \vec{H} = -4\pi \nabla \cdot \vec{m}. \tag{5}$$

This, together with the Landau-Lifshitz equation, leads to a complicated integro-differential equation for the mode solutions. Usually, to calculate these effects the Walker's [5] differential formulation is used and the general solution of this equation is expressed through a fictitious MS-potential function $\psi$: $\vec{H} = -\nabla \psi$. Such a way of solution is used both for continuous-wave FMR [11, 12] and NMR [13] measurements. The question, however, arises: Is the MS-potential wave function $\psi$ really fictitious function?

The power flow density of MS waves propagating along $z$ axis is expressed as

$$P_{MS} = -\frac{i\omega}{4}\left[\psi(\vec{B})^* - (\psi)^*\vec{B}\right]\cdot \vec{e}_z, \tag{6}$$

where $\vec{e}_z$ is the unit vector along $z$ axis. This expression can be obtained by two ways. As it was shown in [14], one derives Eq. (6) from a spectral problem formulation based on quasistatic operator equations for two wave functions: MS-potential function $\psi$ and magnetic flux density $\vec{B}$ ($\vec{B} = -\vec{\mu}(\omega)\nabla\psi$). This operator equation is the following:

$$\hat{L}V = 0, \tag{7}$$

where

$$\hat{L} = \begin{pmatrix} (\vec{\mu})^{-1} & \nabla \\ \nabla \cdot & 0 \end{pmatrix} \tag{8}$$

is the differential-matrix operator and

$$V = \begin{pmatrix} \vec{B} \\ \psi \end{pmatrix} \qquad (9)$$

is the vector function included in the domain of definition of operator $\hat{L}$. In this derivation, no DME are used.

Another derivation is based on use of DME: the power flow density of MS waves formally corresponds to the Poynting vector obtained for the curl electric field and the potential (quasi-magnetostatic) magnetic field [12]. This reveals (together with the McDonald's remarks [8] shown above) a certain physical contradiction. The contradiction becomes evident when one considers the gauge transformation for MS-wave fields derived from the DME. In supposition that there exists a curl electric field $\vec{E}$ defined by the Faraday law, one can introduce a magnetic vector potential: $\vec{E} \equiv -\nabla \times \vec{A}^m$. For monochromatic MS-wave process ($\vec{B} = -\vec{\mu}(\omega) \cdot \nabla \psi$) and based on the Faraday law, we have

$$\nabla^2 \vec{A}^m - \nabla(\nabla \cdot \vec{A}^m) - \frac{i\omega}{c} \vec{\mu}(\omega) \cdot \nabla \psi = 0. \qquad (10)$$

This equation shows that formally two types of gauges are possible. In the first type of a gauge we have:

$$\nabla \cdot \vec{A}^m = 0 \qquad (11)$$

and, therefore,

$$\nabla^2 \vec{A}^m = \frac{i\omega}{c} \vec{\mu}(\omega) \cdot \nabla \psi. \qquad (12)$$

The second type of a gauge is written as

$$\nabla(\nabla \cdot \vec{A}^m) + \frac{i\omega}{c} \vec{\mu}(\omega) \cdot \nabla \psi = 0 \qquad (13)$$

and, therefore,

$$\nabla^2 \vec{A}^m = 0. \qquad (14)$$

The last equation shows that any sources of the electric field are not defined and thus the electric field is not defined at all. So only the first type of a gauge, giving Eq. (12), should be taken into account.

The main point, however, is that the considered above gauge transformation does not fall under the known gauge transformations, neither the Lorentz gauge nor the Coulomb gauge [15], and cannot formally lead to the wave equation. Moreover, to have a wave process one should suppose that there exists a certain physical mechanism describing the effect of transformation of the *curl* ($\vec{E} = -\nabla \times \vec{A}^m$) electric field to the *potential* ($\vec{H} = -\nabla \psi$) magnetic field. From a classical electromagnetic point of view, one does not have such a physical mechanism.

## 3. The gauge electric fields for magnetostatic oscillations

MS oscillations in a one-dimensional linear structure are completely described by scalar wave function $\psi$. In a case of MS wave propagating along $z$ axis in a lossless structure, one has the Schrödinger-like equation [14, 16, 17]:

$$a^{(1)} \frac{\partial^2 \psi(z,t)}{\partial z^2} + a^{(2)} \psi(z,t) = \frac{\partial \psi(z,t)}{\partial t}, \qquad (15)$$

where $a^{(1)}$ and $a^{(2)}$ are imaginary coefficients. Based on this equation one can find the normalized average MS energy of a propagating mode.

The second-order homogeneous differential equation for MS-potential wave function, the Walker equation [5], we write in a form:

$$\hat{G}\psi = 0, \qquad (16)$$

where

$$\hat{G} \equiv -\nabla \cdot (\vec{\vec{\mu}} \nabla) \qquad (17)$$

is a second-order differential operator. Let us represent the MS-potential wave function as a propagating wave in a certain waveguide structure

$$\psi = \tilde{\chi}\, e^{-ikz}, \qquad (18)$$

where $\tilde{\chi}$ is the MS-potential membrane function and $k$ is a propagation constant along $z$ axis. The eigenvalue equation for MS mode $q$ in an axially magnetized ferrite rod is expressed as:

$$\left(\hat{G}_\perp - k_q^2\right)\tilde{\chi}_q = 0, \qquad (19)$$

For a ferrite region we have

$$\hat{G}_\perp = \mu \nabla_\perp^2, \qquad (20)$$

where $\mu$ is a diagonal component of the permeability tensor and $\nabla_\perp^2$ is the two-dimensional (with respect to cross-sectional coordinates) Laplace operator. Outside a ferrite region Eq. (19) becomes the Laplace equation ($\mu = 1$). Double integration by parts on square $S$ – a cross-section of a waveguide structure – of the integral $\int_S (\hat{G}_\perp \tilde{\chi})\,\tilde{\chi}^* dS$ gives the boundary conditions for self-adjointess of operator $\hat{G}_\perp$. For a circular ferrite rod of radius $\Re$ the boundary condition is:

$$\mu\left(\frac{\partial \tilde{\chi}}{\partial r}\right)_{r=\Re^-} - \left(\frac{\partial \tilde{\chi}}{\partial r}\right)_{r=\Re^+} = 0. \qquad (21)$$

or

$$\mu\, (H_r)_{r=\Re^-} - (H_r)_{r=\Re^+} = 0. \qquad (22)$$

MS-potential functions $\tilde{\chi}$ included in the domain of definition of operator $\hat{G}_\perp$ are functions with finite energy. The boundary conditions (21) are called as the essential boundary conditions (EBCs). In accordance with the Ritz method it is sufficient to use basic functions from the energetic functional space with application of the essential boundary conditions [18]. For a constant bias magnetic field, the energy eigenvalue problem for MS waves in a ferrite disk resonator is formulated as the problem defined by the differential equation:

$$\hat{F}_\perp \tilde{\chi}_q = E_q \tilde{\chi}_q \qquad (23)$$

together with the corresponding (essential) boundary conditions [14, 16, 17]. A two-dimensional ("in-plane") differential operator $\hat{F}_\perp$ and energy $E_q$ are determined as:

$$\hat{F}_\perp = \frac{1}{2} g \mu \nabla_\perp^2, \qquad (24)$$

$$E_q = \frac{1}{2} g k_q^2, \qquad (25)$$

where $g$ is the unit dimensional coefficient. The energy orthonormality in a ferrite disk described as

$$(E_q - E_{q'}) \int_S \tilde{\chi}_q \tilde{\chi}_{q'}^* dS = 0, \qquad (26)$$

acquires now a real physical meaning. There are the Hilbert functional space of MS-potential functions $\tilde{\chi}$. Because of discrete energy eigenstates of MS-wave oscillations resulting from structural confinement in a case of a normally magnetized ferrite disk, one can consider the oscillating system as a collective motion of quasiparticles – the light magnons [16, 17]. The energy eigenvalue problem formulated based on the EBCs shows that a ferrite disk with magnetic-dipolar-mode oscillations is a Hamiltonian system.

The considered above spectral problem bears, however, a formal character. The essential boundary conditions differ from the physical situation demanding continuity for normal components of $\vec{B}$ as the boundary conditions. The last ones (called as natural boundary conditions (NBCs) [18]) are necessarily satisfied by the boundary conditions of functions $V$ – the functions included in the domain of definition of operator $\hat{L}$ – but not by functions with finite energy. Evidently, the equation $\nabla \cdot \vec{B} = 0$ is satisfied for the NBCs, but not for the EBCs.

For the NBC problem described by Eqs. (7), we represent function $V$ for propagating waves as

$$V = \tilde{V} e^{-i\kappa z}, \tag{27}$$

where tilder means MS-wave membrane functions: $\tilde{V} = \begin{pmatrix} \vec{\tilde{B}} \\ \tilde{\varphi} \end{pmatrix}$, $\kappa$ is a propagation constant along $z$ axis. The eigenvalue equation for MS mode $m$ is expressed as:

$$\left( \hat{L}_\perp - i\kappa_m \hat{R} \right) \tilde{V}_m = 0, \tag{28}$$

where

$$\hat{R} \equiv \begin{pmatrix} 0 & \vec{e}_z \\ -\vec{e}_z & 0 \end{pmatrix}, \tag{29}$$

subscript $\perp$ means differentiation over a waveguide cross section. Integration by parts on $S$ – a square of an open MS-wave waveguide – of the integral $\int_S (\hat{L}_\perp \tilde{V}) \tilde{V}^* dS$ gives the contour integral in a form $\oint_C (\tilde{B}_r \tilde{\varphi}^* - \tilde{B}_r^* \tilde{\varphi}) dC$, where $C$ is a contour surrounding a cylindrical ferrite core and $B_r$ is a component of a membrane function of the magnetic flux density normal to contour $C$. Operator $\hat{L}_\perp$ becomes self-adjoint for homogeneous boundary conditions (continuity of $\tilde{\varphi}$ and $\tilde{B}_r$) on contour $C$. Based on the homogeneous boundary conditions one obtains the orthogonality relation:

$$(\kappa_m - \kappa_n) \int_S \left( \hat{R} \tilde{V}_m \right) \left( \tilde{V}_n \right)^* dS = 0. \tag{30}$$

Formulation of the NBC spectral problem is based on the homogeneous boundary conditions for the radial component of $\vec{B}$ which is described as

$$\mu (H_r)_{r=\Re^-} - (H_r)_{r=\Re^+} = -i\mu_a (H_\theta)_{r=\Re^-}. \tag{31}$$

Here $H_r$ and $H_\theta$ are, respectively, radial and azimuth components of the RF magnetic field and $\mu_a$ is the off-diagonal component of tensor $\vec{\mu}$. For magnetostatic solutions: $H_r = -\dfrac{\partial \psi}{\partial r}$ and $H_\theta = -\frac{1}{r}\frac{\partial \psi}{\partial \theta}$. Because of a cylindrical symmetry of a sample, the membrane function $\tilde{\varphi}$ is written as $\tilde{\varphi} = \tilde{\varphi}(r)\tilde{\varphi}(\theta)$. With a formal supposition that an angular part is described as $\tilde{\varphi}(\theta) = e^{-i\nu\theta}$, one rewrites Eq. (31) as:

$$\mu\left(\frac{\partial \tilde{\varphi}}{\partial r}\right)_{r=\mathfrak{R}^-} - \left(\frac{\partial \tilde{\varphi}}{\partial r}\right)_{r=\mathfrak{R}^+} = -\frac{\mu_a}{\mathfrak{R}}\nu(\tilde{\varphi})_{r=\mathfrak{R}^-}. \tag{32}$$

From this formal representation, it becomes evident that for a given sign of $\mu_a$, the solutions for MS-wave functions depend on a sign of $\nu$. For an axially magnetized ferrite rod this fact was shown, for the first time, by Joseph and Schlömann [19]. So because of the boundary conditions we have different functions $\tilde{\varphi}$ for positive and negative directions of an angle coordinate when $0 \le \theta \le 2\pi$. It means that functions $\tilde{\varphi}$ cannot be considered as the single-valued functions.

The fact that solution of the boundary problem is dependent on a sign of $\nu$, rises a question about validity of the energy orthogonality relation for MS-wave modes. For a system for which a total Hamiltonian is conserved, there should be single valuedness for egenfunctions [20]. Since the eigenstates of Eq. (28) are not single valued, one should find a phase factor that will make the states single valued.

Following a standard way of solving boundary problems in mathematical physics [18,21], let us consider two joint boundary problems: the main boundary problem and the conjugate boundary problem. The problems are described by differential equations which are similar to Eq. (28). The main problem is expressed by a differential equation:

$$\left(\hat{L}_\perp - i\beta\,\hat{R}\right)\tilde{V} = 0. \tag{33}$$

The conjugate problem is expressed by an equation:

$$\left(\hat{L}_\perp^\circ - i\beta^\circ\,\hat{R}\right)\tilde{V}^\circ = 0. \tag{34}$$

From a formal point of view, it is supposed initially that these are different equations: there are different differential operators, different eigenfunctions and different eigenvalues. A form of differential operator $\hat{L}_\perp^\circ$ one gets from integration by parts:

$$\int_S (\hat{L}_\perp \tilde{V})(\tilde{V}^\circ)^* dS = \int_S \tilde{V}(\hat{L}_\perp^\circ \tilde{V}^\circ)^* dS + \oint_C P(\tilde{V}, \tilde{V}^\circ) dC, \qquad (35)$$

where $P(\tilde{V}, \tilde{V}^\circ)$ is a bilinear form. For an open ferrite structure [a core ferrite region (*F*) is surrounded by a dielectric region (*D*)] the homogeneous boundary conditions for functions $\tilde{V}$ and $\tilde{V}^\circ$ give

$$\oint_C [P^{(F)}(\tilde{V}, \tilde{V}^\circ) + P^{(D)}(\tilde{V}, \tilde{V}^\circ)] dC = 0. \qquad (36)$$

In this case operator $\hat{L}_\perp$ is a self-conjugate operator. For self-conjugate operators, the orthogonality relations can be derived. When one considers functions $\tilde{V}$ and $\tilde{V}^\circ$ as the fields of modes *m* and *n*, one obtains the orthogonality relation (30).

We demand continuity of $\tilde{\varphi}$ and $\tilde{B}_r$ on the border *C*. So the boundary condition (36) we should write as

$$\oint_C \{[(\tilde{B}_r)_{r=\Re^-} - (\tilde{B}_r)_{r=\Re^+}](\tilde{\varphi}^\circ)^*_{r=\Re} - (\tilde{\varphi})_{r=\Re}[(\tilde{B}_r^\circ)_{r=\Re^-} - (\tilde{B}_r^\circ)_{r=\Re^+}]^*\} dC = 0. \qquad (37)$$

We now uncover the expression for magnetic flux density $\vec{B}$ in Eq. (37). Since in a ferrite region $\tilde{B}_r = \mu \frac{\partial \tilde{\varphi}}{\partial r} + i\mu_a \frac{\partial \tilde{\varphi}}{\partial \theta}$ and in a dielectric $\tilde{B}_r = \frac{\partial \tilde{\varphi}}{\partial r}$, one has

$$\oint_C [\mu\left(\frac{\partial \tilde{\varphi}}{\partial r}\right)_{r=\Re^-} - \left(\frac{\partial \tilde{\varphi}}{\partial r}\right)_{r=\Re^+}](\tilde{\varphi}^\circ)^*_{r=\Re} - (\tilde{\varphi})_{r=\Re}[\mu\left(\frac{\partial \tilde{\varphi}^\circ}{\partial r}\right)_{r=\Re^-} - \left(\frac{\partial \tilde{\varphi}^\circ}{\partial r}\right)_{r=\Re^+}]^* dC +$$

$$\oint_C [(i\mu_a \frac{\partial \tilde{\varphi}}{\partial \theta})(\tilde{\varphi}^\circ)^* - (\tilde{\varphi})(i\mu_a \frac{\partial \tilde{\varphi}^\circ}{\partial \theta})^*]_{r=\Re} dC = 0. \qquad (38)$$

In the above equation we represented a contour integral (37) as a sum of two contour integrals.

For a case of the single-valuedness, the first integral in Eq. (38) should be equal to zero [see Eq. (21)]. Since, however, functions $\tilde{\varphi}$ are not single-valued functions, the first integral in Eq. (38) is not equal to zero.

Let us introduce a new membrane function $\tilde{\eta}$:

$$\tilde{\eta}(\rho,\alpha) = \begin{cases} \gamma_+ \tilde{\varphi}_+ \\ \gamma_- \tilde{\varphi}_- \end{cases}, \tag{39}$$

where

$$\gamma_\pm = a_\pm e^{-iq_\pm \theta}. \tag{40}$$

The function $\tilde{\varphi}$ changes a sign when $\theta$ is rotated by $2\pi$. Therefore, in order to cancel this sign change, $\gamma_\pm$ must change its sign to preserve the single-valued nature of $\tilde{\varphi}$. From this we conclude that $e^{-iq_\pm 2\pi} = -1$. That is

$$q_\pm = l\frac{1}{2}, \tag{41}$$

where $l = \pm 1, \pm 3, \pm 5, ...$.

Now we rewrite Eq. (39) as follows:

$$\tilde{\varphi}_\pm = \frac{1}{\gamma_\pm}\tilde{\eta} = \delta_\mp \tilde{\eta}, \tag{42}$$

where

$$\delta_\mp = \frac{1}{a_\pm}e^{-iq_\mp \theta} \equiv f_\mp e^{-iq_\mp \theta}. \tag{43}$$

In the above relations, evidently, $a_+ = -a_-$ and $f_+ = -f_-$. To have proper normalization we will take $|a_\pm| = |f_\mp| = 1$.

We substitute expression (42) into Eq. (38). Since the boundary conditions for single-valued functions $\tilde{\eta}$ should correspond to the boundary conditions (21), the first integral in Eq (38) becomes equal to zero. With use of substitution (42) we have from Eq. (38)

$$\oint_C \{[i\frac{\partial(\delta_\mp\tilde{\eta})}{\partial\theta}](\delta_\mp^\circ\tilde{\eta}^\circ)^* - (\delta_\mp\tilde{\eta})[i\frac{\partial(\delta_\mp^\circ\tilde{\eta}^\circ)}{\partial\theta}]^*\}_{r=\Re}dC = 0. \tag{44}$$

This gives:

$$\int_0^{2\pi}\{[\delta_\mp(\delta_\mp^\circ)^*][(i\frac{\partial\tilde{\eta}}{\partial\theta})(\tilde{\eta}^\circ)^* - (\tilde{\eta})(i\frac{\partial\tilde{\eta}^\circ}{\partial\theta})^*]\}_{r=\Re}d\theta + $$
$$\int_0^{2\pi}\{[\tilde{\eta}(\tilde{\eta}^\circ)^*][(i\frac{\partial\delta_\mp}{\partial\theta})(\delta_\mp^\circ)^* - (\delta_\mp)(i\frac{\partial\delta_\mp^\circ}{\partial\theta})^*]\}_{r=\Re}d\theta = 0. \tag{45}$$

Both functions, $\delta_\mp$ and $\eta$, describe a periodic process with respect to angle $\theta$. Let us introduce a generalized periodic function $y$ and consider the eigenvalue equation

$$i\frac{\partial}{\partial\theta}y = uy, \tag{46}$$

where $u$ is a real quantity. We introduce now a problem with eigenvalue equation conjugate to Eq. (46) (with eigen function $y^\circ$ and eigenvalue $u^\circ$) and consider an integral $\int_0^{2\pi}(i\frac{\partial y}{\partial\theta})(y^\circ)^*d\theta$. Using integration by parts of this integral, one finds that when $u$ (and $u^\circ$) are integer numbers (including 0): $u = 0, \pm 1, \pm 2, \pm 3,...$ and when $u$ (and $u^\circ$) are half-integer numbers: $u = \pm\frac{1}{2}, \pm\frac{3}{2}, \pm\frac{5}{2},...$, just only in these two separate cases operator $\hat{J}_z \equiv i\frac{\partial}{\partial\theta}$ is a self-conjugate operator and one can write the orthogonality relation:

$$(u - u^\circ)\int_0^{2\pi}y(y^\circ)^*d\theta = 0. \tag{47}$$

For any mixed situation (when, for example, $u$ is an integer number and $u^\circ$ is a half-integer number), functions $y$ and $y^\circ$ are not mutually orthogonal. It means that the spectral problems for integer egenvalues $u$ should be considered separately from the spectral problem for half-integer eigenvalues $u$. Based on this consideration of the orthogonality relation for generalized functions $y$, one should conclude that $\delta_\mp (\delta_\mp^\circ)^* = 1$ in the first integral of Eq. (45), while $\tilde{\eta}(\tilde{\eta}^\circ)^* = 1$ in the second integral of Eq. (45).

The first integral in Eq. (45) is evidently equal to zero. So, as a result, one has from Eq. (45):

$$\int_0^{2\pi} \{[(i\frac{\partial \delta_\mp}{\partial \theta})(\delta_\mp^\circ)^* - (\delta_\mp)(i\frac{\partial \delta_\mp^\circ}{\partial \theta})^*]\}_{r=\Re} d\theta = 0. \tag{48}$$

The transformation (39) restores the single valuedness, but now there is a nonzero vector-potential-type term:

$$A_\theta^m \equiv \int_0^{2\pi} [(i\frac{\partial \delta_\mp}{\partial \theta})(\delta_\mp^\circ)^*]_{r=\Re} d\theta = q_\mp. \tag{49}$$

Since $q_\mp = \mp\frac{1}{2}, \mp\frac{3}{2}, \mp\frac{5}{2},...$, there are the *positive and negative vector-potential-type terms*. The superscript "$m$" means that there is the *magnetic* vector-potential-type term.

Function $\tilde{\eta}$ is identical to function $\tilde{\chi}$. The confinement effect for magnetic-dipolar oscillations requires proper phase relationships to guarantee single-valuedness of the wave functions. To compensate for sign ambiguities and thus to make wave functions single valued we added a vector-potential-type term to the MS-potential Hamiltonian. This procedure is similar to the procedure made by Mead for the Born-Oppenheimer wave functions [22, 23]. The corresponding flux of pseudo-electric field $\vec{\in}$ (the gauge field) through a circle of radius $\Re$ is obtained analogously to [22]:

$$\int_S \vec{\in} \cdot d\vec{S} = \oint_C \vec{A}_\theta^m \cdot d\vec{C} = \Phi^e, \tag{50}$$

where $\Phi^e$ is the flux of pseudo-electric field. The energy levels are periodic in the electric flux $\Phi^e$. There should be the positive and negative fluxes. These different-sign fluxes should be inequivalent to avoid the cancellation.

Similar to electromagnetic theory, the vector potential $\vec{A}_\theta^m$ is defined up to a gauge transformation. By performing the formal transformation

$$\tilde{\varphi}' = \tilde{\varphi}\, e^{ip(\theta)}, \tag{51}$$

it is easy to show that

$$(A_\theta^m)' = \vec{A}_\theta^m + \nabla_\theta p(\theta). \tag{52}$$

Despite the fact that $\oint_C \vec{A}_\theta^m \cdot d\vec{C} \neq 0$, one has $\nabla \times \vec{A}_\theta^m = 0$. So the gauge electric field $\in$ is not related to the Faraday-law electric field $\vec{E}$.

## 4. Persistent magnetic currents in magnetic-oscillation disks

The value $\oint_C \vec{A}_\theta^m \cdot d\vec{C} \neq 0$ can be observable. The above analysis of a phase factor that makes the states single valued and so makes a total Hamiltonian to be conserved is related to a topological effect in a closed system. In this case the results are gauge invariant and the Stokes theorem can be used.

In such a closed system, there should be a certain internal mechanism which creates a non-zero vector potential $\vec{A}_\theta^m$. This internal mechanism becomes evident when one compares the EBC (providing single-valuedness) described by Eq. (21) and the NBC (not providing single-valuedness) described by Eq. (31). The difference arises from the term in the right-hand side, which contains the gyrotropy parameter (the off-diagonal component of the permeability tensor, $\mu_a$) and the annular magnetic field $\vec{H}_\theta$. Just due to this term a non-zero vector potential appears. The annual magnetic field $\vec{H}_\theta$ is a singular field existing only in an infinitesimally narrow cylindrical layer abutting (from a ferrite side) to a border of a ferrite disk. One does not have any special conditions connecting radial and azimuth components of magnetic fields on other (inner or outer) circular contours, except contour *C*. Because of such an annual magnetic field, the notion of an effective circular magnetic current can be considered.

Let us formally introduce a quantity of a magnetic current:

$$\vec{j}^m(z) \equiv \frac{1}{4\pi} i\omega\mu_a \vec{H}_\theta(z). \tag{53}$$

We can rewrite the boundary condition (31) as follows:

$$\delta(r-\Re)\left[\frac{1}{4\pi}\omega\mu(H_r)_{r=\Re^-} - \frac{1}{4\pi}\omega(H_r)_{r=\Re^+}\right] = -i^m, \tag{54}$$

where $i^m$ is a density of an effective boundary magnetic current defined as

$$\vec{i}^m(z) \equiv \delta(r-\Re)\frac{1}{4\pi}i\omega\mu_a(\vec{H}_\theta(z))_{r=\Re^-} = \delta(r-\Re)\vec{j}^m(z). \tag{55}$$

In supposition that membrane ("flat") functions $\tilde{\chi}$ form a complete basis in the energy functional space with use of boundary condition (21), it becomes evident that the effective boundary magnetic current slips from the main properties of this functional space. This current, being a persistent magnetic current, cannot be considered as a single-valued function.

A singular "border" MS-potential function $\delta_\mp$ is described by Eq. (43). For a certain MS oscillating mode in a ferrite disk we can represent an annual magnetic field as

$$(H_\theta(z))_{r=\Re^-} = -\xi(z)\nabla_\theta \delta_\mp = -\xi(z)\frac{1}{\Re}\frac{\partial \delta_\mp}{\partial \theta}\bigg|_{r=\Re^-}, \tag{56}$$

where function $\xi(z)$ describes *z*-distribution of the MS potential in a ferrite disk [14]. For a circular effective boundary magnetic current we have now:

$$\left(i^{m}(z)\right)=-\xi(z)\frac{i\omega\mu_{a}}{4\pi\Re}\frac{\partial\delta_{\mp}}{\partial\theta}\bigg|_{r=\Re^{-}}=-\xi(z)\frac{\omega\mu_{a}q_{\mp}f_{\mp}}{4\pi\Re}e^{-iq_{\mp}\theta}. \tag{57}$$

The "border" MS-potential functions $\delta_{\mp}$, being characterized by the "spin coordinates", are antisymmetrical functions. At the same time, as it follows from Expr. (57), the effective magnetic currents are described by symmetrical functions with respect to the "spin coordinates". In other words, the effective magnetic current has the same direction for the "right" and "left" spinning states. The signs of magnetic current $i^{m}$ are different for different signs of $\mu_{a}$. However, the "positive" ($\mu_{a}>0$) and "negative" ($\mu_{a}<0$) magnetic currents do not mutually compensate each other since for different signs of $\mu_{a}$ we have structures with different symmetries. This will be clear from further consideration.

Circulation of current $i^{m}$ along contour $C$ gives a nonzero quantity when $q_{\mp}$ is a number divisible by $\frac{1}{2}$:

$$D(z)=\oint_{C}(i^{m})\,dC=\Re\int_{0}^{2\pi}(i^{m})\,d\theta=i\xi(z)\frac{\omega\mu_{a}}{2\pi}f_{\mp}. \tag{58}$$

The fields existing inside a ferrite disk form a very special field structure outside a disk. For nonzero circulation $D(z)$ one can formally define an electric moment of a whole ferrite disk resonator (in a region far away from a disk) as follows [24]:

$$a^{e}=-i\frac{1}{2c}\int_{0}^{h}dz\oint_{C}(\vec{\rho}\times\vec{i}^{\,m})\cdot\vec{e}_{z}\,dC=\frac{\omega\mu_{a}}{4\pi c}\Re f_{\mp}\int_{0}^{h}\xi(z)dz, \tag{59}$$

where $h$ is a disk thickness.

In the above consideration, transport round a closed path (which gives Berry's phase factor $\pi$) is the excursion of the system in time. The circular motion described by "border" MS-potential functions $\delta_{\mp}$ is the time-reversal-odd process. Also $\mu_{a}$ is the time-reversal-odd function and, therefore, an electric moment $a^e$ should be the time-reversal-even function. At the same time, since magnetic current $\vec{i}^{\,m}$ is an axial vector, it follows that vector $\vec{\rho} \times \vec{i}^{\,m}$ is a polar vector. So an electric moment $\vec{a}^e$ is the parity-odd time-reversal-even function.

## 5. Symmetry properties of oscillating magnetic modes

Self-induced quasistationary magnetic fields are characterized by dynamical symmetry breaking. Let us introduce the quantity $Q \equiv \mu_a f_\mp$. One can distinguish the case when $Q > 0$ and the case when $Q < 0$. This discriminates, in fact, two situations: (a) directions of a circular transport and magnetic current $\vec{i}^{\,m}$ are the same and (b) directions of a circular transport and magnetic current $\vec{i}^{\,m}$ are opposite.

The energy eigenstate (see Eq. (25)) is determined by two waves propagating in a ferrite disk: the forward and backward waves with respect to axis $z$ [17]. Since in a normally magnetized disk the forward and backward waves propagate along opposite directions of a bias magnetic field, they are the time-reversal-odd waves. These waves are characterized by different signs of $\mu_a$ [11]. The circular motion described by "border" MS-potential functions is the time-reversal-odd process as well. Evidently, for a given energy eigenstate there should be the same sign of $Q$ (and, therefore, the same direction an electric moment $\vec{a}^e$) for the forward and backward waves. At the same time, the direction of the "spinning rotation" with respect to the direction of a polar vector $\vec{a}^e$ is different for the forward and backward waves. So one has different symmetry properties of the forward and backward waves. To a certain extent, this resembles the "particle – antiparticle" symmetry properties in elementary particle physics. The above analysis gives an evidence for four types of oscillating modes: two different-symmetry (forward and backward) modes for $Q > 0$ and two different-symmetry (forward and backward) modes for $Q < 0$.

As it follows from the theoretical analysis [14, 16, 17] and experimental studies [6, 7, 25], the energy levels of oscillating modes in a normally magnetized ferrite disk are distinguished by discrete quantities of a bias magnetic field $H_0$. For zero magnetic field $H_0$, the modes with $Q > 0$ and $Q < 0$ are degenerate with respect to the energy. When a bias magnetic field is applied, different orientations of an electric moment $\vec{a}^e$ (parallel or antiparallel with respect to $\vec{H}_0$) correspond to different energy levels. So one may have the energy splitting between two cases: $\vec{a}^e \cdot \vec{H}_0 > 0$ and $\vec{a}^e \cdot \vec{H}_0 < 0$. Such energy splittings (which we can characterize as the *magnetoelectric* energy splittings) were experimentally observed in [25], when a normally magnetized ferrite disk was placed in a maximum the electric component of a cavity field.

## 6. Conclusion

The problem of the self-induced quasielectrostatic and quasimagnetostatic fields is especially important in understanding mechanisms of interaction of small temporally-dispersive-material samples with electromagnetic radiation. In particular, electrostatic resonances of isolated nanoparticles have recently attracted substantial interest because of intriguing possibility of obtaining very strong and localized electric fields. However, when the theory predicts multiresonance electrostatic (plasmon) oscillations in small temporally-dispersive-permittivity samples [2, 26, 27], experiments of the electromagnetic response [28] show, in fact, only a very few absorption peaks. Contrarily, in a case of small temporally-dispersive-permeability disks one can find (both from the theory [14, 16, 17] and experiments [6, 7, 25]) the pictures of multiresonance magnetostatic oscillations. The present paper gives further explanation of these phenomena. Our main standpoint is that, unlike the known results of the self-induced quasielectrostatic fields, the self-induced quasimagnetostatic fields in small magnetic samples are the Hilbert-space modes. Together with interaction with the magnetic component of the electromagnetic radiation, a small magnetic disk interacts with the electric component of the electromagnetic field. This is because of dynamical symmetry properties of magnetostatic modes. The dynamical symmetry breaking in quasistatic magnetic oscillations shows special-type gauge transformation for the fields.